\documentclass[twocolumn,prb,showpacs,superscriptaddress,amsmath,amssymb]{revtex4-1}
\usepackage{graphicx}
\usepackage{dcolumn}
\usepackage{color}
\usepackage{bm}
\newcommand{\br}{\mathbf{r}}

\newcommand{\hh}{\hat{H}}
\newcommand{\ha}{\hat{a}}

\newcommand{\he}{\mathcal{H}}

\begin{document}

\title{Spin Density Waves and Domain Wall Interactions in Nanowires}

\author{N. Sedlmayr}
\email{sedlmayr@physik.uni-kl.de}
\affiliation{Department of Physics, University of Kaiserslautern,
D-67663 Kaiserslautern, Germany}
\author{V. K. Dugaev}
\affiliation{Department of Physics, Rzesz\'ow University of Technology,
al. Powsta\'nc\'ow Warszawy 6, 35-959 Rzesz\'ow, Poland}
\affiliation{Department of Physics and CFIF, Instituto Superior T\'ecnico,
TU Lisbon, av. Rovisco Pais, 1049-001, Lisbon, Portugal}
\author{J. Berakdar}
\affiliation{Department of Physics, Martin-Luther-Universit\"at Halle-Wittenberg,
Heinrich-Damerow-Str.~4, 06120, Halle, Germany}

\date{\today}

\begin{abstract}
We investigated how the dynamics of a domain wall are affected by the presence of spin density waves in a ferromagnetic wire. Domain walls and other scattering centres can cause coherent spin density waves to propagate through a wire when a current is applied. In some cases the spin torque due to these scattered electrons can be enhanced such that it is on a par with the exchange and anisotropy energies controlling the shape and dynamics of the domain wall. In such a case we find that the spin density waves enhance the current induced domain wall motion, allowing for domain wall motion with smaller current pulses. Here we consider a system involving two domain walls and focus on how the motion of the second domain wall is modified by the spin density waves caused by the presence of the first domain wall.
\end{abstract}


\maketitle

\section{Introduction}

The dynamics of a domain wall (DW) in a ferromagnetic wire has received much interest due to both its fundamental physical importance, but also in light of  possible future applications.\cite{yamanouchi04,yamaguchi04,parkin08,Marrows2005} Coupling between the carriers of spin polarized currents and the magnetic moments forming DW alters not only the transport properties\cite{ebels00,chopra02,ruster03} of the wire but also the magnetization itself. The current traveling through the wire couples to the domain wall causing it to move through the wire. Thus, differently orientated collinear regions of the wire may serve as discrete bits that can be steered  by the current. In order to increase the efficiency of these devices\cite{parkin08,parkin10} it is necessary to increase the density of the DWs. Therefore even a weak interaction between DWs can become important.
This coupling can also distort the DW during its propagation through the wire. Some examples from the now voluminous literature can be found here.\cite{Bertotti2007285,Thomas06,Marrows2005}

So far most work has concentrated on how the spin polarized currents will affect the domain walls. However, there are also feedback processes which can lead to interactions between the DWs and spin density fluctuations. The relevant effects are enhanced considerably when considering low dimensional structures.\cite{klaui08} An effective RKKY interaction between two domain walls in a wire can be mediated by the current electrons.\cite{sedlmayr09,Sedlmayr20101419,sedlmayrpssb} The energy profile of this interaction then favours particular alignments and positions of the domain walls. Furthermore any change in the spin density caused by the presence of one domain wall can have an effect on a wall further down the wire, defined in the direction of electron flow.

For applications it is most interesting to consider a high density of DWs in a wire. In this case it is crucial
to inspect how  the DWs affect each other when a spin polarized current is sent down the wire.
In a previous article\cite{sedlmayr09} we looked at  how the current mediated RKKY-like interaction changes the magnetization dynamics with a focus on the motion of relatively sharp walls. These walls were  treated as localized moments. Here we extend this idea to consider how a long (adiabatic) domain wall is distorted in the presence of a spin density wave caused by scattering from a previous domain wall. The spin density wave acts as an effective applied magnetic field. However, because the spin torque of the current electrons is no longer uniform, it is not a homogeneous field. This will tend to distort the shape of the domain walls.

Furthermore this non-uniform spin density  allows the domain wall to be set into motion using a smaller current pulse than would otherwise be the case. This is closely related to its effect as an effective magnetic field. This offers an alternative possible solution\cite{PhysRevLett.105.217203} to the problems of overheating associated with sending too large a current through the nanowires. Though we note that in order to obtain a spin density of the appropriate order of magnitude one must consider very narrow wires, of a few atoms across.

It is now possible to fabricate ferromagnetic wires composed of single atomic sites on a lattice.\cite{Gambardella,PhysRevB.56.2340,PhysRevLett.73.898,PhysRevB.57.R677} Surprisingly these wires show both ferromagnetic order and contain regions of non-collinearity,\cite{RevModPhys.81.1495} in other words: domain walls. In the low dimensional limit of these metallic wires
 the effect of the spin density corrections on the domain wall may become on the order of that of the exchange and magnetic anisotropy.
 This can cause severe distortion of the domain wall profile and alter the way it travels through the wire. We note that the spin density corrections must not of necessity originate from a DW, any spin coherent scattering centre will lead to the behaviour described in this article, for example a magnetic impurity or region of non-collinearity in an otherwise single domain ferromagnet. Although here we concentrate on the case of a sharp domain wall as the scatterer, the generalization to other scenarios is straightforward. The spin density for a specific material can be varied by two parameters, the overall and relative magnitudes of the two spin channels. By varying these parameters we can consider a variety of scatterers.

Our set up thus consists of a ferromagnetic wire with two domain walls present. A current is then applied to the wire. We focus on how the motion of the second domain wall (defined with respect to the current direction) is affected by the spin density waves caused by the presence of the first domain wall in the wire.

\section{Theoretical formulation}

We start with a  zero temperature general Hamiltonian for a nanowire describing non-interacting conduction electrons of spin $\alpha$, $\ha_\alpha^\dagger(\br)$, coupled with a strength $J$ to some non-uniform bulk magnetization, $\vec{M}(\br)$:\cite{ll9,Blundell}
\begin{eqnarray}
\hh'&=&\int
d\br\sum_\alpha\ha_\alpha^\dagger(\br)[\hat{\xi}\delta{\alpha\beta}-J\vec{\sigma}_{\alpha\beta}.\vec{M}(\br)]\ha_\beta(\br).
\end{eqnarray}
$\hat{\xi}$ is the kinetic energy operator. The inhomogeneity can be partially dealt with by a local vector gauge
transformation.\cite{kor,tatara,dugaev1} After this transformation we will have a uniform Zeeman
splitting term and a spin-dependent spatially varying potential,
$U_{\alpha\beta}(\br)$, which describes the scattering from any non-collinear configurations of the magnetization. This transformation describes a local rotation in spin-space to align the magnetization direction throughout the wire. It is possible for any form of $\vec{M}(\br)$ which has a constant magnitude. The gauge transformation is
\begin{eqnarray}\label{gauge}
\begin{pmatrix}
\ha^{\textrm{old}}_1(\br)\\
\ha^{\textrm{old}}_2(\br)\end{pmatrix}
=\mathbf{T}(\br)
\begin{pmatrix}\ha^{\textrm{new}}_1(\br)\\
\ha^{\textrm{new}}_2(\br)\end{pmatrix}
\end{eqnarray}
defined such that ${\bm T}^\dag(\br)\, \vec{{\bm \sigma}} \cdot \vec{n}(\br)\, {\bm T}(\br)
=\sigma ^z$, where $\vec{n}$ is the unit vector along $\vec{M}$.\footnote{
In this work
we will consider the transverse dynamics of the domain wall by setting $\vec{M}=M\vec{n}$
where $\vec{n}$ is a unit vector field and is the dynamical variable here. The longitudinal
dynamics in $\vec{M}$  occurs at higher energies and is not considered here.
The local coupling constant $JM$ is material and carrier-type dependent and varies
in a considerable range; e.g. we estimate it to be $1.7$ eV for Co and $2.24$ eV for Fe.
On the other hand, if a localized magnetic structure is considered a scattering center instead of  DW, say
a single Mn$_{12}$ molecular magnet (H. B. Heersche \emph{et al.}, Phys. Rev. Lett. \textbf{96}, 206801 (2006); M. H. Jo \emph{et al.}, Nano Lett. \textbf{6}, 2014 (2006))  we estimate   $JM= 1$meV (see R.-Q. Wang \emph{et al.}, Phys. Rev. Lett. \textbf{105}, 057202 (2010) for the exchange coupling).} Our new Hamiltonian is then
\begin{eqnarray}
\hh=\int\label{hamiltonian}
d\br\sum_{\alpha\beta}\ha_\alpha^\dagger(\br)[\hat{\xi}\delta_{\alpha\beta}
-JM\sigma^z_{\alpha\beta}+U_{\alpha\beta}(\br)]\ha_\beta(\br).
\end{eqnarray}
with the scattering potential given by
\begin{eqnarray}
\mathbf{U}(\br)=-\frac{1}{2m}[2\vec{\mathbf{A}}(\br).\partial_{\br}+
\partial_\br.\vec{\mathbf{A}}+\vec{\mathbf{A}}^2(\br)],
\end{eqnarray}
and $\vec{{\bm A}}(\br)={\bm T}^\dag (\br)\, \nabla _\br\, {\bm T}(\br)$ is a gauge potential.

For the case of two domain walls in a nanowire with the first DW located at $z=0$ and the second DW at $z=z_0$, and of widths $L'$ and $L$ respectively, we have
\begin{eqnarray}\label{magnetization2}
\vec{M}(z)=M\begin{pmatrix}\cos[\theta(z)]\sin[\varphi(z)]\\ \sin[\theta(z)]\sin[\varphi(z)]\\ \cos[\varphi(z)]
\end{pmatrix}
\end{eqnarray}
where
\begin{eqnarray}
\varphi(z)=\underbrace{\pi-\cos^{-1}\big[\tanh[z/L']\big]}_{=\varphi_1(z)}\nonumber\\+\underbrace{\pi-\cos^{-1}\big[\tanh[(z-z_0)/L]\big]}_{=\varphi_2(z)}.
\end{eqnarray}
The angle $\theta(z)$ is used to give the walls a different orientation. Around the first wall we set it arbitrarily to $0$ and around the second
to an angle $\theta_0$, which therefore defines the relative orientation of the two walls. Cross terms between the walls are very small and can be neglected provided $z_0\gg L,L'$, allowing us to approximate the scattering potential,
\begin{eqnarray}
\mathbf{U}(z)\approx\mathbf{U}_1(z)+\mathbf{U}_2(z),
\end{eqnarray}
as independent contributions from each DW. To see some interaction effect between the DWs we must nonetheless of course have the distance between the DWs as less than the spin coherence lengthscale in the system. Now
\begin{eqnarray}\label{u1}
\mathbf{U}_1(z)&=&\mathbb{I}\frac{[\varphi_1'(z)]^2}{8m}+i\bm{\sigma}^y
\bigg[\frac{\varphi_1''(z)}{4m}+\frac{\varphi_1'(z)\partial_z}{2m}\bigg]\textrm{ and}\\
\mathbf{U}_2(z)&=&\mathbb{I}\frac{[\varphi_2'(z)]^2}{8m}+i\bm{\sigma}^y
\bigg[\frac{\varphi_2''(z)}{4m}+\frac{\varphi_2'(z)\partial_z}{2m}\bigg]\cos(\theta_0)\nonumber\\&&-i\bm{\sigma}^x
\bigg[\frac{\varphi_2''(z)}{4m}+\frac{\varphi_2'(z)\partial_z}{2m}\bigg]\sin(\theta_0).\label{u2}
\end{eqnarray}
If $L>\lambda_F$ then we can also neglect the $\varphi_2''(z)$ and $[\varphi_2'(z)]^2$ terms which are of order of $(\lambda_F/L)^2$.

\begin{figure}
\includegraphics[width=0.45\textwidth]{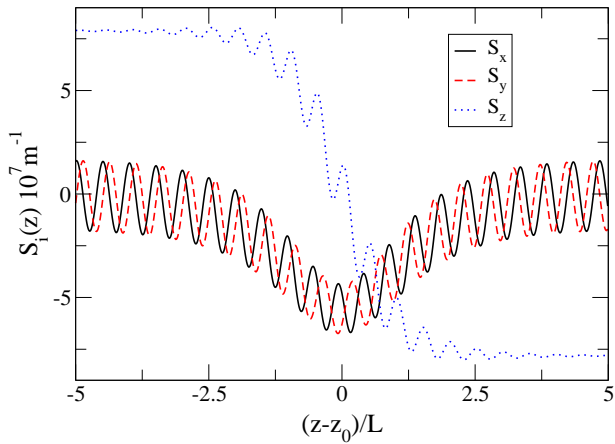}
\caption[Spin Density]{(Color online) The spin density of the carrier electrons around the second domain wall, centred at $Z_0$ and measured in carrier spin per unit length. One can clearly see both the spin density waves and the overall profile of the second domain wall.
The solid (black) curve is for $S_x(z)$, the dashed (red) is for $S_y(z)$,
and the dotted (blue) is for $S_z(z)$.}
\label{spindensity}
\end{figure}

The correction to the spin-density of a single scattered wave of spin $\delta$ is
\begin{eqnarray}\label{spindensityeqn}
\Delta\vec{S}_{\delta}&=&\bm{\delta\psi}^\dagger_{\varepsilon\delta}\vec{\bm{\sigma}}\bm{\delta\psi}_{\varepsilon\delta}.
\end{eqnarray}
$\bm{\delta\psi}_{\varepsilon\delta}$ is the scattered wavefunction, from an incoming particle of spin $\delta$, at an energy $\varepsilon$. It can be calculated within the Born approximation provided that $L'\gtrsim\lambda_F$.\cite{sedlmayr09} This tells us the nature of the spin waves in our system. These spin density corrections result in an inter-DW spin torque and RKKY-like interaction, details of which can be found in Sedlmayr et al.\cite{sedlmayr09,Sedlmayr20101419,sedlmayrpssb}. The result for the spin density around the second domain wall at an arbitrarily chosen configuration, $\theta_0=\pi/4$, is shown in figure \ref{spindensity}. Plotted for $\lambda_F=0.367$nm, $L'=\lambda_F$, $L=10\lambda_F$, and $JM=2.24$eV.
The adiabatic change in spin density across the width of the second DW is clearly visible, and is modified by the presence of the spin density waves. These oscillations in the spin density will tend to inhomogeneously distort the domain wall profile away from its equilibrium profile.

\section{Motion of the second domain wall}\label{dwmotion}

We wish to find the effect of the scattered electrons on the second domain wall. We take the spin density of the scattered electrons and calculate its effect on the second DW using the now standard Landau-Lifschitz-Gilbert (LLG) equation approach.\cite{ll9,gilbert04,zhang04}  In our case this gives us
\begin{eqnarray}\label{llg}
\partial_t\vec{M}(z,t)=-\frac{1}{1+\alpha^2}\bigg[\gamma \vec{M}(z,t)\times\he\qquad\qquad\nonumber\\+\frac{\alpha\gamma}{M}\vec{M}(z,t)\times(\vec{M}(z,t) \times\he)\nonumber\\
+\frac{b_j(1+\alpha\xi)}{M^2}\vec{M}(z,t)\times(\vec{M}(z,t)\times(\vec{j}.\nabla)\vec{M}(z,t))\nonumber\\+ \frac{b_j(\xi-\alpha)}{M}\vec{M}(z,t)\times(\vec{j}.\nabla)\vec{M}(z,t)\bigg].
\end{eqnarray}
$\gamma$ is the gyromagnetic ratio:
$\gamma=\frac{g\mu_B}{\hbar}$,
where $g$ is the Land\'e factor, and $\alpha$ is a phenomenological constant characterizing the magnetic damping of the system. The constants for the non-adiabatic terms are: $\vec{j}$ is the current density; $\xi=\tau_{ex}/\tau_{sf}$ is the ratio between the exchange and spin-flip time scales; and
$b_j=\frac{P\mu_B}{eM(1+\xi^2)}$,
with $P$ the polarization.
The effective field is
\begin{eqnarray}\label{sizes}
\he(z)=J\vec{S}(z)+\alpha_{\textrm{ex}}\frac{\partial^2\vec{M}(z)}{\partial z^2}+KM_x(z)\vec{x}.
\end{eqnarray}
$\alpha_{\textrm{ex}}$ is the exchange coupling. The coefficient $K$ characterizes the magnetic anisotropy in the wire. We have chosen here $x$ as the anisotropy axis. We use\cite{ll9,Blundell} $\alpha_{\textrm{ex}}=J/Ma$ and $L_2=\pi\sqrt{K/\alpha_{\textrm{ex}}}$. ($a$ is the lattice spacing.) Note however that the domain wall width can be changed by geometric effects as well as by changing the anisotropy or the exchange energy. The effective field $\he$, which the DW experiences is composed of the spin density fluctuations of the carriers, the exchange interaction of the DW itself, and the magnetic anisotropy of the system. The exchange interaction tends to oppose distortions on the wire and will impose an upper limit on how much the DW can be distorted by a given spin density.

We calculate the dynamics resulting from equation \eqref{llg} using the boundary conditions $\vec{M}(0,z)=\vec{M}_0(z)$ and $\vec{M}(t,z_0\pm z_b)=\vec{M}_0(z\pm z_b)$, where $\vec{M}_0(z)$ is the magnetization of equation \eqref{magnetization2}. $z_b$ must be taken to be large enough to insert no artifacts in the results.
Note that from equation \eqref{sizes} it is clear that for a sizeable effect we require $aL^2>|S|$, where $S$ is the spin density. This can be reached by considering the experimentally realizable limit of a chain of atoms with a cross section of approximately $\sigma_{cs}=\lambda_F^2$. The lattice spacing is taken as $a=1$~nm and we use here the Fermi wavelength of iron: $\lambda_F=0.367$~nm. Also we take $M=1.72\times10^6$~A$\textrm{m}^{-1}$, $JM=2.24$~eV, $j_e=-2.33\times 10^8$~Am~${}^{-2}$, and $\xi=0.011$.\cite{zhang04} The second domain wall has a width of $L=10\lambda_F$, the first of the order of the Fermi wavelength, $L'=\lambda_F$. The DW we consider the motion of is therefore relatively narrow. This gives us the advantage of making our results clearer. However it could also limit the applicability of the model. We note though that a DW would be narrower in a wire with a small cross section than in the bulk or thin film case, meaning that
upon patterning of the wire the DW width can be decreased. Nonetheless the results here would hold for DW widths of the order of $10$nm. The case for DWs of larger widths is discussed below. Finally: $\alpha=0.01$, and the exchange and anisotropy are as defined above ($x$ is therefore a hard magnetic axis). The second DW's configuration is begun at an arbitrary angle, $\theta_0=\pi/4$. Altering the initial angle of the second DW does not significantly affect the current induced motion of the second domain wall. The speed of the DW's motion is the same irrespective of the value of $\theta_0$. Qualitatively the same motion is seen except for some small distortions in the DW profile which depend on $\theta_0$.

\begin{figure*}
\includegraphics[width=0.8\textwidth]{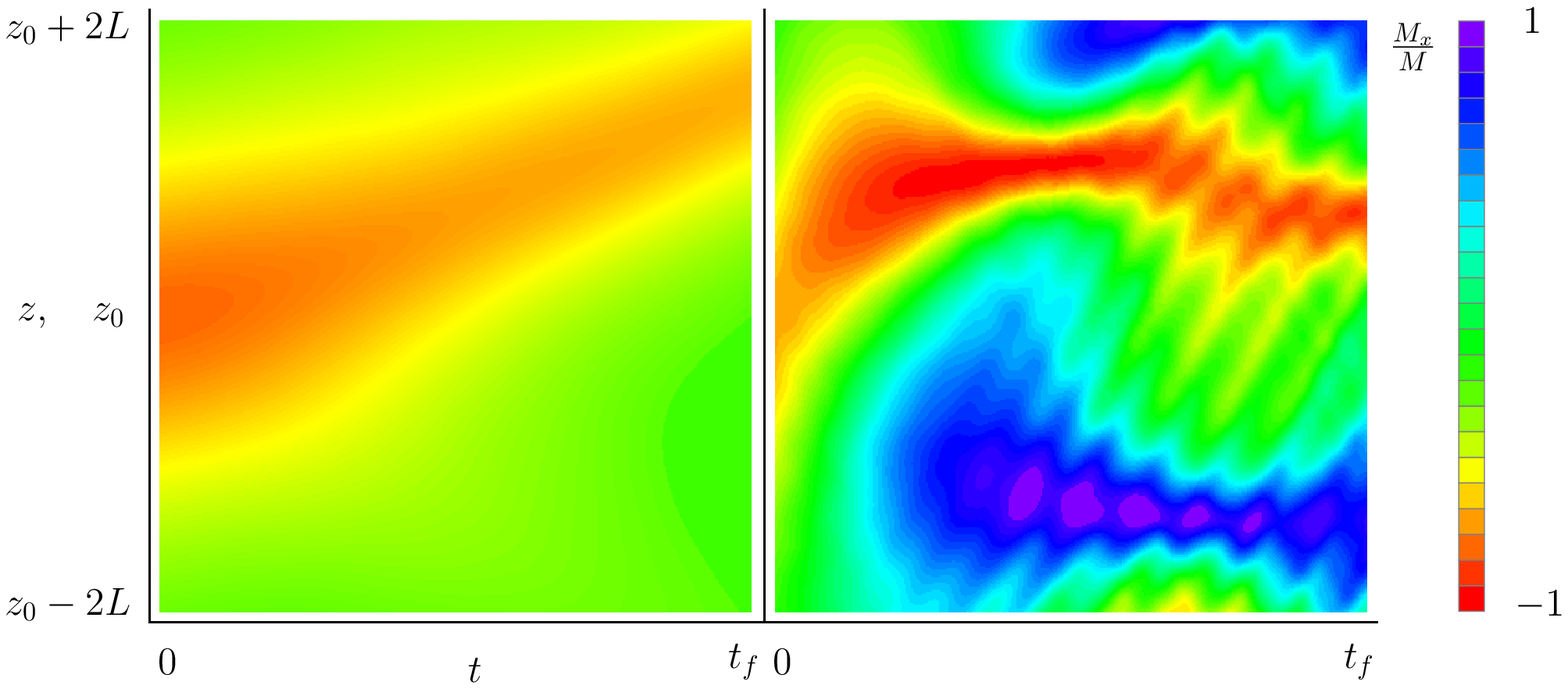}
\caption[$M_x(t,z)$: Magnetization Dynamics]
{(Color online) A contour plot of the $x$-component of the magnetization dynamics $M_x(t,z)$ around the second wall \emph{without} an applied current, centered at $z_0$. The left hand figure is without the spin density corrections, the right hand figure includes the spin density corrections. The left hand figure shows motion caused by the anisotropy in the wire, on the right hand side the spin density waves cause additional distortions. $t_f=2.69\times10^{-13}$s.}\label{dwmotionx}
\end{figure*}

\begin{figure*}
\includegraphics[width=0.8\textwidth]{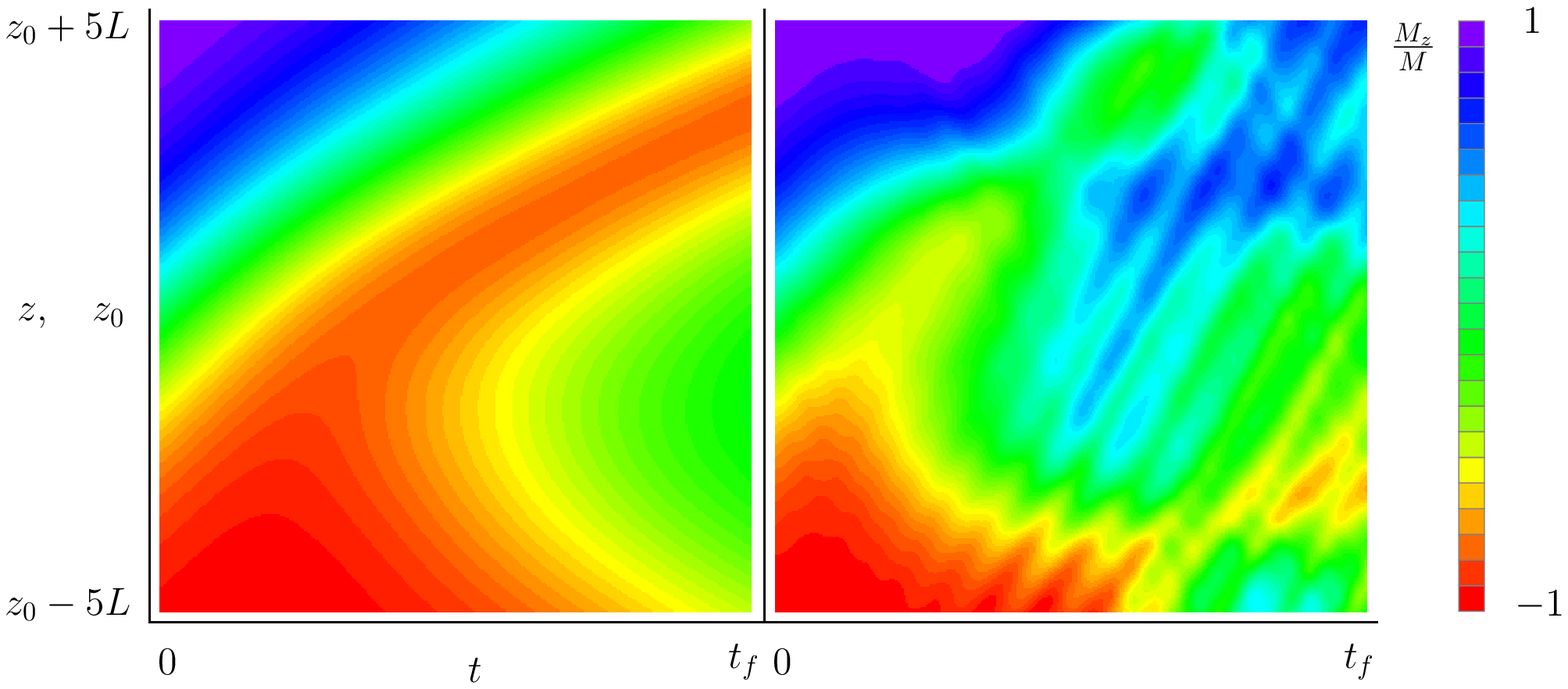}
\caption[$M_z(t,z)$: Magnetization Dynamics]
{(Color online) A contour plot of the $z$-component of the magnetization dynamics $M_z(t,z)$ around the second wall \emph{without} an applied current, centered at $z_0$. The left hand figure is without the spin density corrections, the right hand figure includes the spin density corrections. The left hand figure shows motion caused by the anisotropy in the wire, on the right hand side the spin density waves cause additional distortions. One can see that the presence of the spin density readily distorts the DW profile. $t_f=2.69\times10^{-13}$s.}\label{dwmotionz}
\end{figure*}

Firstly let us consider the effects of the spin density wave without considering the current induced motion in equation \eqref{llg}, plotted in figures \ref{dwmotionx} and \ref{dwmotionz}. In the right hand side (RHS) of figures \ref{dwmotionx} and \ref{dwmotionz} over a short period of time the wall is distorted into a low energy position, as in the case without the spin density corrections (the left hand side (LHS) of figures \ref{dwmotionx} and \ref{dwmotionz}), but as the spin density fluctuates on a scale much shorter than the second wall we see no coherent evolution of the domain wall over a longer period of time. The motion seen when there is no spin density correction present, the LHS of figures \ref{dwmotionx} and \ref{dwmotionz}, is because we have not started the walls in their equilibrium position with respect to the anisotropy in the system. In this case the domain wall attempts to find its lowest energy configuration.
What is most clear from these figures is that the spin density is able to completely distort the profile of the domain wall. Rather than being able to shift to its equilibrium configuration the domain wall is severely distorted by the spin density torque. Its low energy position is now a compromise between not just the anisotropy and exchange energies, but also the spin density torque exerted on the wall. The low energy profile with respect to the spin density contributions would be an oscillating shape, but this is opposed by the exchange energy cost it brings.

\begin{figure*}
\includegraphics[width=0.8\textwidth]{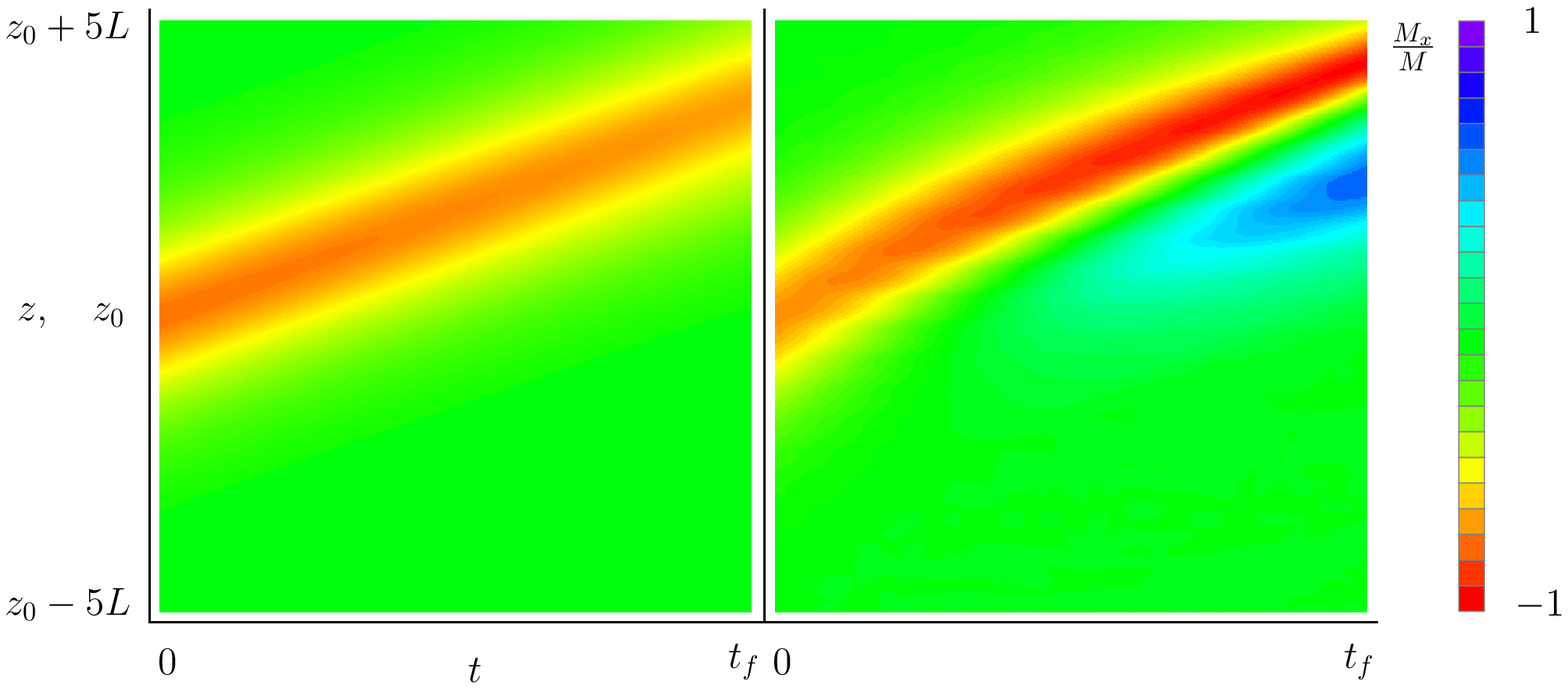}
\caption[$M_x(t,z)$: Magnetization Dynamics]
{(Color online) A contour plot of the $x$-component of the magnetization dynamics $M_x(t,z)$ around the second wall, centered at $z_0$. This includes the current terms for the magnetization dynamics, with $j_e=-2.33\times 10^8$~Am~${}^{-2}$. The left hand figure is without the spin density corrections, the right hand figure is the full result, showing the accelerated start up to the DW motion due to the spin density terms. $t_f=6.73\times10^{-14}$s.}\label{dwmcurrentx}
\end{figure*}

Now let us consider what happens when we include the current terms in equation \eqref{llg}.  The case without any correction to the spin density from scattering from a first domain wall is shown in the LHS of figure \ref{dwmcurrentx}, here we focus on the $x$-component of the magnetization. One simply sees the usual current induced motion of the domain wall. The RHS  of figure \ref{dwmcurrentx} includes the spin density corrections. After some time one also sees a distortion of the domain wall profile caused by the spin density corrections. The new domain wall profile then undergoes coherent motion with this new shape.
The presence of the spin density wave also has a small but noticeable transient effect at the beginning of the domain wall's motion, the wall is set into motion more quickly than in the case without the spin density wave. This effect is short lived and the long term motion is, in this case, similar to without the spin density oscillations. After a period of time some small distortion of the DW profile does appear. However, distortions similar to those found when the current terms do not contribute are not seen.

\begin{figure*}
\includegraphics[width=0.8\textwidth]{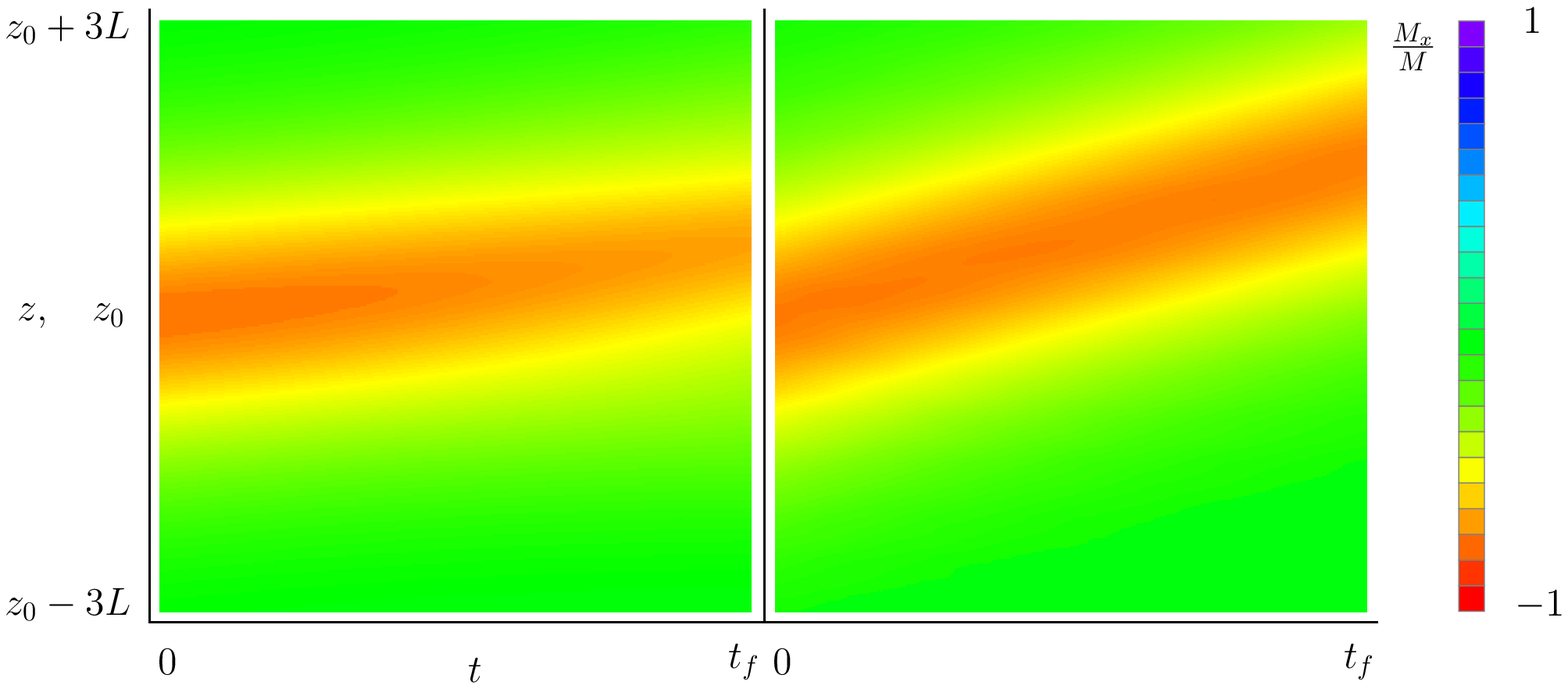}
\caption[$M_x(t,z)$: Magnetization Dynamics]
{(Color online) A contour plot of the $x$-component of the magnetization dynamics $M_x(t,z)$ around the second wall, centered at $z_0$. The left hand figure is without the spin density corrections, the right hand figure is the full result including spin density corrections. In this case we apply a smaller current: $j_e=-5.82\times 10^7$~Am~${}^{-2}$. The quicker motion of the domain wall due to the spin density correction is now clearly visible in the right hand figure as compared to the left hand figure. $t_f=6.73\times10^{-14}$s.}\label{dwmsmallj}
\end{figure*}

If we lower the applied current density we find the effects more pronounced. In fact the presence of the spin density appreciably speeds up the motion of the DW when a small current is applied, see figure \ref{dwmsmallj}. This opens up the possibility that in such systems the presence of closely packed multiple domain walls assists in the current induced motion of subsequent DWs, allowing them to be shifted with a smaller applied current pulse.
The spin density wave acts as a local magnetic field, and in principle gives one more parameters with which to control the current induced domain wall motion. Namely the overall and relative amplitudes of the spin density correction channels, equation \eqref{spindensityeqn}.

\begin{figure}
\includegraphics[width=0.45\textwidth]{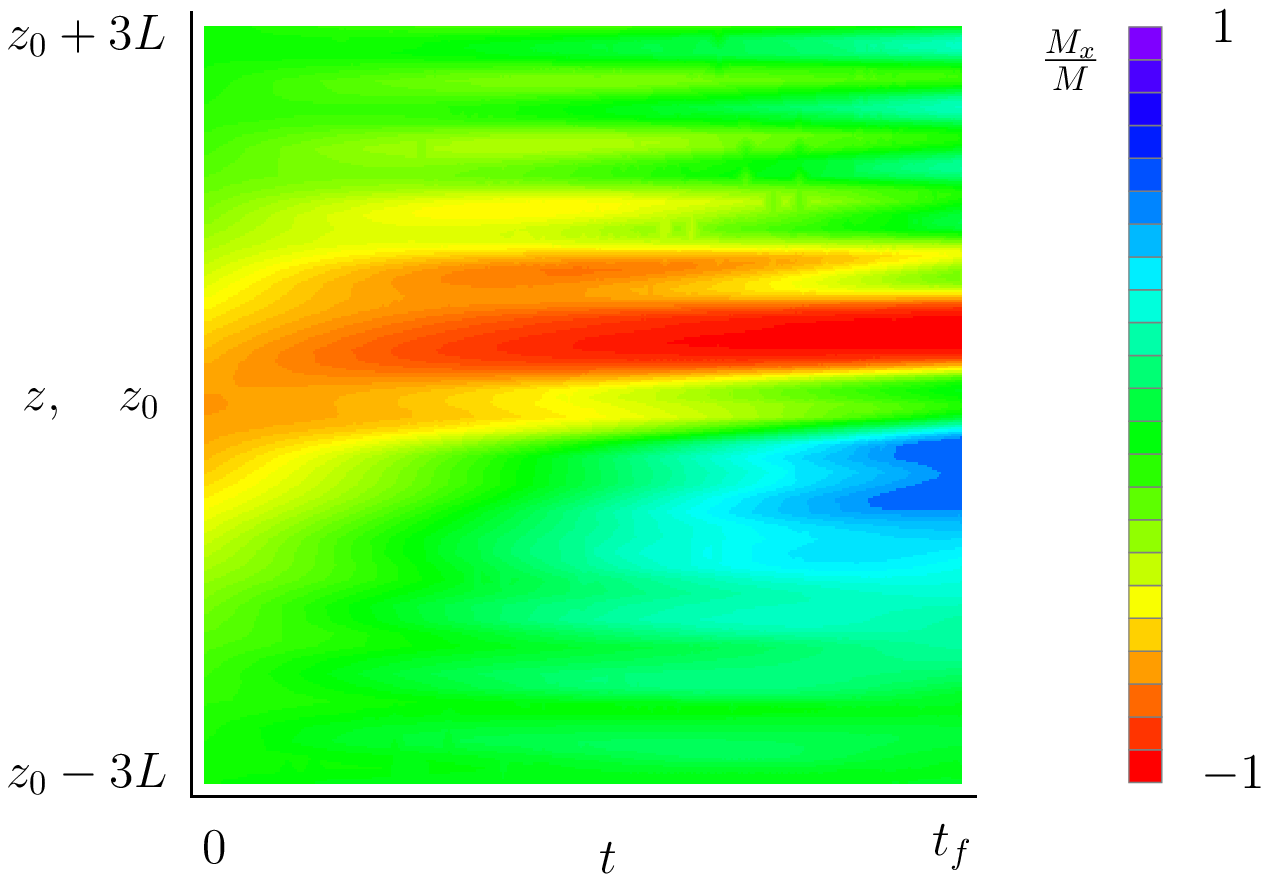}
\caption[$M_x(t,z)$: Magnetization Dynamics]
{(Color online) A contour plot of the $x$-component of the magnetization dynamics $M_x(t,z)$ around the second wall, centered at $z_0$. Here we consider a longer DW of width $L=40$nm. The initial current induced motion of the DW is still enhanced by the spin density. for longer times thechange in DW velocity is negligible. What is seen is a narrowing of the DW width as it moves caused by the spin density wave. The applied current is $j_e=-2.33\times 10^8$~Am~${}^{-2}$. $t_f=6.73\times10^{-14}$s.}\label{longdw}
\end{figure}

We can also consider what happens for DWs of larger widths. For a DW of width $L=10$nm the main result is still valid. The spin density wave aids the current induced DW motion. However if we increase the DW width to $L=40$nm these effects are already negligible for longer timescales. The reason for this is clear, for a longer DW the spin density wave is oscillating too much over its extension  to have a coherent effect. Nonetheless for short timescales one still sees quicker initial motion of the DW. Furthermore one can see a gradual squeezing of the DW caused by the spin density wave, see Fig.~\ref{longdw}. The edges of the DW are gradually squeezed as they attempt to precess in the effective magnetic field of the rapidly oscillating spin density wave. This process is limited by the increase in exchange energy for a narrower DW. The DW ends with a width of approximately $10-20$nm undergoing current induced motion. This effect is of course absent without the spin density wave.

\section{Discussion and conclusions}

We have investigated how spin density waves caused by spin polarized currents scattering from domain walls affects the domain wall's magnetization dynamics. In the one dimensional limit of atomic chains the spin-torque exerted by the spin density waves can be sizable, comparable even with the exchange energy of the bulk ferromagnetic system. For pinned domain walls the inhomogeneous spin distorts the domain wall profile into a new shape which minimizes the relevant energies of the domain wall. However, when the domain wall is free to move through the system the walls are distorted only slightly. In this case the main effect of the spin waves is to propel the domain wall into motion quicker than in their absence.

The relatively strong effect of the current-induced spin density wave on the motion
of the domain wall is similar to the effect of an external magnetic field.
This is because the magnetization related to the spin wave appears in the equation of
motion as an external field, and, on the other hand, it is
known that by using magnetic field one can easily put the domain wall into motion.
However, in the case of a spin density wave, the effective field acting on the
domain wall is not homogeneous. On the contrary, it is strongly oscillating.
Correspondingly, it can affect the short-range interaction of magnetic
moments making unstable the usual shape of the domain wall. Being strongly inhomogeneous,
the spin-wave induces some displacements and certain disorder of the wall,
which is important for the DW motion because the first displacement allows
depinning of the wall to put it into motion.

In this connection, we propose to combine the spin-wave-induced dynamics
with the short pulses of current. It can be realized with a combination of
dc current, generating the spin density wave, and strong current pulses.

\section*{Acknowledgments}
This work is partly supported by FCT Grant No.~PTDC/FIS/70843/2006
in Portugal, by the DFG contract BE 2161/5-1, and by the Graduate School of MAINZ (MATCOR).

\bibliography{referencesnonadiabatictorque}

\end{document}